\documentclass[a4paper,fleqn,usenatbib]{mnras}

\usepackage{newtxtext,newtxmath}
\usepackage[T1]{fontenc}
\usepackage{ae,aecompl}

\usepackage{natbib}

\usepackage{graphicx}	
\usepackage{amsmath}
\usepackage{hyperref}
\usepackage{listings}
\usepackage{color}
\usepackage{float}
\usepackage{xspace} 
\usepackage{url}

\newcommand{\kepler}{\textsl{Kepler}\xspace}
\newcommand{\tablenotemark}[1]{$^{#1}$}
\title[Stellar Activity with Gaia]{Spotting Stellar Activity Cycles in Gaia Astrometry}

\author[Morris et al.]{
Brett M. Morris,$^{1}$\thanks{E-mail: bmmorris@uw.edu}
Eric Agol$^{1}$
James. R. A. Davenport,$^{2, 3}$
Suzanne L. Hawley$^{1}$
\\
$^{1}$Astronomy Department, University of Washington, Seattle, WA 98195, USA\\
$^{2}$Department of Physics \& Astronomy, Western Washington University, 516 High St., Bellingham, WA 98225, USA\\
$^{3}$NSF Astronomy and Astrophysics Postdoctoral Fellow
}

\date{Accepted 2018 February 23. Received 2018 February 22; in original form 2017 December 24}

\pubyear{2018}

\begin{document}
\label{firstpage}
\pagerange{\pageref{firstpage}--\pageref{lastpage}}
\maketitle

\begin{abstract}
Astrometry from Gaia will measure the positions of stellar photometric centroids to unprecedented precision. We show that the precision of Gaia astrometry is sufficient to detect starspot-induced centroid jitter for nearby stars in the Tycho-Gaia Astrometric Solution (TGAS) sample with magnetic activity similar to the young G-star KIC 7174505 or the active M4 dwarf GJ 1243, but is insufficient to measure centroid jitter for stars with Sun-like spot distributions. We simulate Gaia observations of stars with 10 year activity cycles to search for evidence of activity cycles, and find that Gaia astrometry alone likely can not detect activity cycles for stars in the TGAS sample, even if they have spot distributions like KIC 7174505. We review the activity of the nearby low-mass stars in the TGAS sample for which we anticipate significant detections of spot-induced jitter.
\end{abstract}

\begin{keywords}
astrometry --- Sun: activity --- stars: activity, low-mass, starspots, individual: GJ 1243, KIC 7174505, AX Mic, $\sigma$ Dra, GX And, HD 79211, LHS 3531, HD 222237, HD 36395, Gl 625
\end{keywords}

\section{Introduction}

The ESA Gaia mission will accurately measure the astrometric positions of many of the nearest stars in the Milky Way \citep{Gaia2016b}. With time-resolved astrometry from Gaia, it will be possible to detect the reflex motions of stars in binary or multi-star systems, and stars in systems with massive planets.



One source of noise that will inflate the observed scatter in astrometric measurements is stellar activity. Stars like the Sun have starspot covering fractions of order 0.03\% in the optical. The distribution of dark spots on the stellar surface changes as the star rotates and as starspots evolve, so the center of light or centroid of a star will vary with time in the optical. The effects of stellar surface inhomogeneities on Gaia astrometry has been considered as a source of noise in planet searches for main sequence stars \citep{Eriksson2007, Catanzarite2008, Lanza2008}, and in parallax measurements of red supergiants \citep{Chiavassa2011}. 

We consider here the potential for detecting magnetic activity cycles from the apparent astrometric shifts of stars due to starspots. The solar activity cycle (as it would be observed in astrometric jitter) lasts about 11 years \citep[see review by][]{Hathaway2015}, which is similar to the extended mission duration of Gaia. Many of the nearest Gaia astrometry targets are low mass main sequence stars, and those with observed activity cycles have periods from a few to $\sim 10$ years \citep{GomesdaSilva2012, Robertson2013, Mascareno2016}. 

We estimate the starspot-induced astrometric jitter produced by the starspots of stars with well-constrained spot asymmetries, the Sun, the active M4 dwarf GJ 1243, and the young G-star KIC 7174505 in Section~\ref{sec:sim}. We compute the anticipated scale of the spot-induced astrometric jitter for nearby main sequence stars in the Gaia sample, and compare the jitter to Gaia's anticipated astrometric precision in Section~\ref{sec:jitter}. We conclude with a brief review of the literature on the stellar activity of the most promising Gaia targets in Section~\ref{sec:targets}.

\section{Centroid estimating algorithm}

We approximate the stellar centroid for a star with non-overlapping circular spots using either an analytic or numerical approximation. Here we briefly outline each technique, and validation between the two techniques. An implementation of these algorithms in Python is available online\footnote{\url{https://github.com/bmorris3/mrspoc}}. 

\subsection{Analytic centroid approximation}
If all spots are smaller than $R_{\mathrm{spot}}/R_\star = 0.1$ we apply the analytic approximation to compute the stellar centroids. We first integrate the total stellar flux of the unspotted, limb-darkened star,
\begin{equation}
F_{\star, \mathrm{unspotted}} = \int_{0}^{R_\star} 2 \pi r \, I(r) dr,
\end{equation}
where $I(r)$ is a quadratic limb-darkening law and $r$ is in units of
angle, so that $2\pi rdr$ is solid angle. For all stars, we use the solar limb-darkening parameters.

We define cartesian sky-plane coordinates $(x,y)$, with the origin placed at the center of the star, $\hat{x}$  aligned with the stellar equator, and $\hat{y}$ aligned with the stellar rotation axis. We describe each starspot with an ellipse with centroid ${\bf r}_i = (x_i, y_i)$, and $r_i = \vert {\bf r}_i \vert$.  We can compute the (negative) flux contribution from each spot by computing the approximate spot area and contrast. A circular spot will be foreshortened near the stellar limb. The foreshortened circular spot can be approximated with an ellipse with semi-major axis $R_{\mathrm{spot}}$ and semi-minor axis $R_{\mathrm{spot}} \sqrt{1 - (r_i/R_\star)^2}$. 

Since these spots are small compared to the stellar radius ($R_{\mathrm{spot}}/R_\star < 0.1$), we adopt one limb-darkened contrast for the entire spot, $c_{ld} = (1-c) I(r)$, where $c$ is the flux contrast in the spot relative to the photosphere flux. We discuss reasonable values of $c$ in Section~\ref{sec:contrast}.

The integrated spot flux is 
\begin{equation}
F_{\mathrm{spot}, i} = - \pi R_{\mathrm{spot}}^2 c_{ld} \sqrt{1 - (r_i/R_\star)^2}, 
\end{equation}
and accounting for all $N$ spots, the position of the stellar centroid $(x_c, y_c)$ is  
\begin{eqnarray}
x_c &=& \left(\sum_{i=1}^{N} x_iF_{\mathrm{spot}, i}\right) / F_{\star, \mathrm{spotted}}\\
y_c &=& \left(\sum_{i=1}^{N} y_iF_{\mathrm{spot}, i}\right) / F_{\star, \mathrm{spotted}},
\end{eqnarray}
where the spotted flux of the star is
\begin{equation}
F_{\star, \mathrm{spotted}} = F_{\star, \mathrm{unspotted}}  + \sum_{i=1}^{N} F_{\mathrm{spot}, i}.
\end{equation}
This approximation is valid for spots that are small compared to the stellar radius, or small compared to the scale of limb-darkening variation across the stellar disk.

\subsection{Numerical centroid approximation}

For a spot configuration with large spots $R_{\mathrm{spot}}/R_\star > 0.1$, we compute the stellar centroid with a simple numerical approximation. We create a square grid with 3000 pixels on a side, and calculate the flux within each pixel given the quadratic limb darkening law for the star, $I(r)$.  We define the boundaries of foreshortened spots using the same geometric approximation as in the previous section, but this time we multiply all pixels within the spot by the spot contrast $c$. 

This numerical method is more computationally expensive than the analytic method, but it does not assume one limb-darkened contrast for the entire spot, and thus it is a better approximation for large spots.

\subsection{Validation}

\begin{figure}
\centering
\includegraphics[scale=0.75]{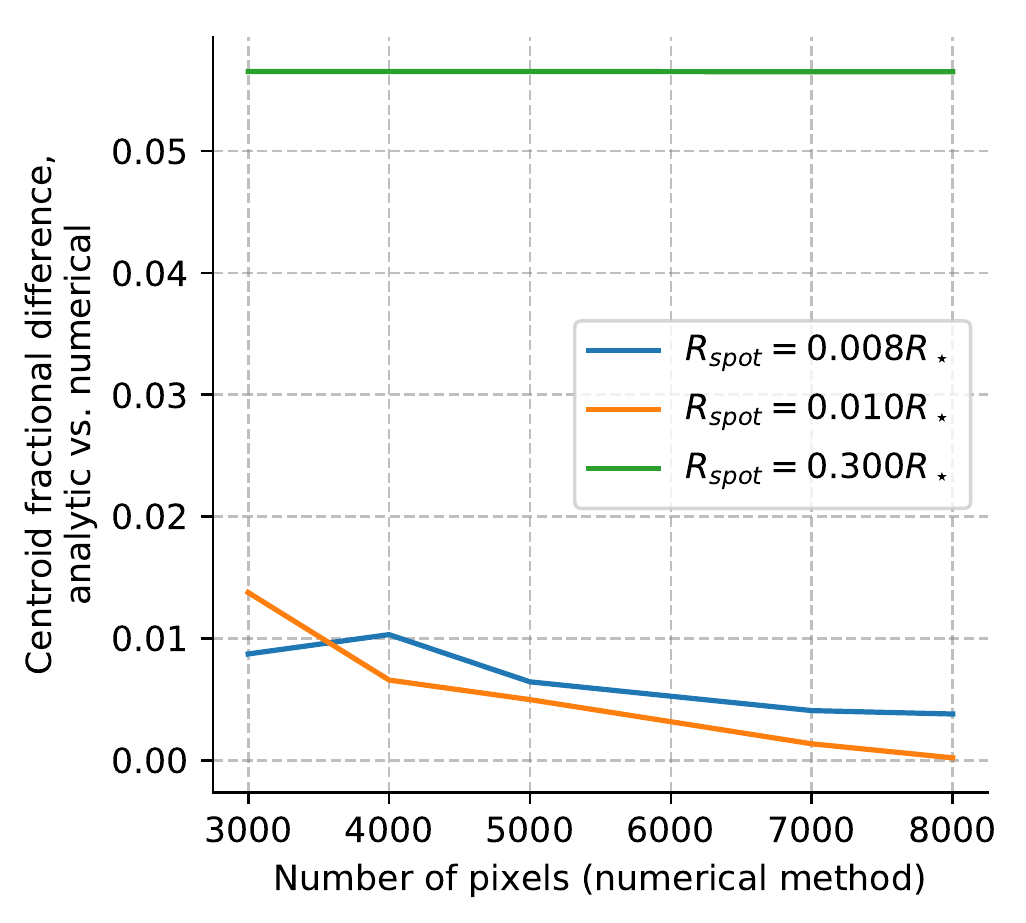}
\caption{We validate the two centroid approximations against one another, finding that the stellar centroids agree to better than $6\%$ for all spots considered in this work.}
\label{fig:valid}
\end{figure}

We confirm that the numerical and analytic approximations produce similar results by computing the stellar centroids for example spot configurations with both methods. The maximum fractional difference between centroids for spots of various sizes, and for numerical approximations with different numbers of pixels, in Figure~\ref{fig:valid}. We find that the centroid agreement between methods is better than 6\% for pixel grids with 3000 pixels on a side or more.

\subsection{Starspot Contrasts} \label{sec:contrast}

\begin{figure}
\centering
\includegraphics[scale=0.85]{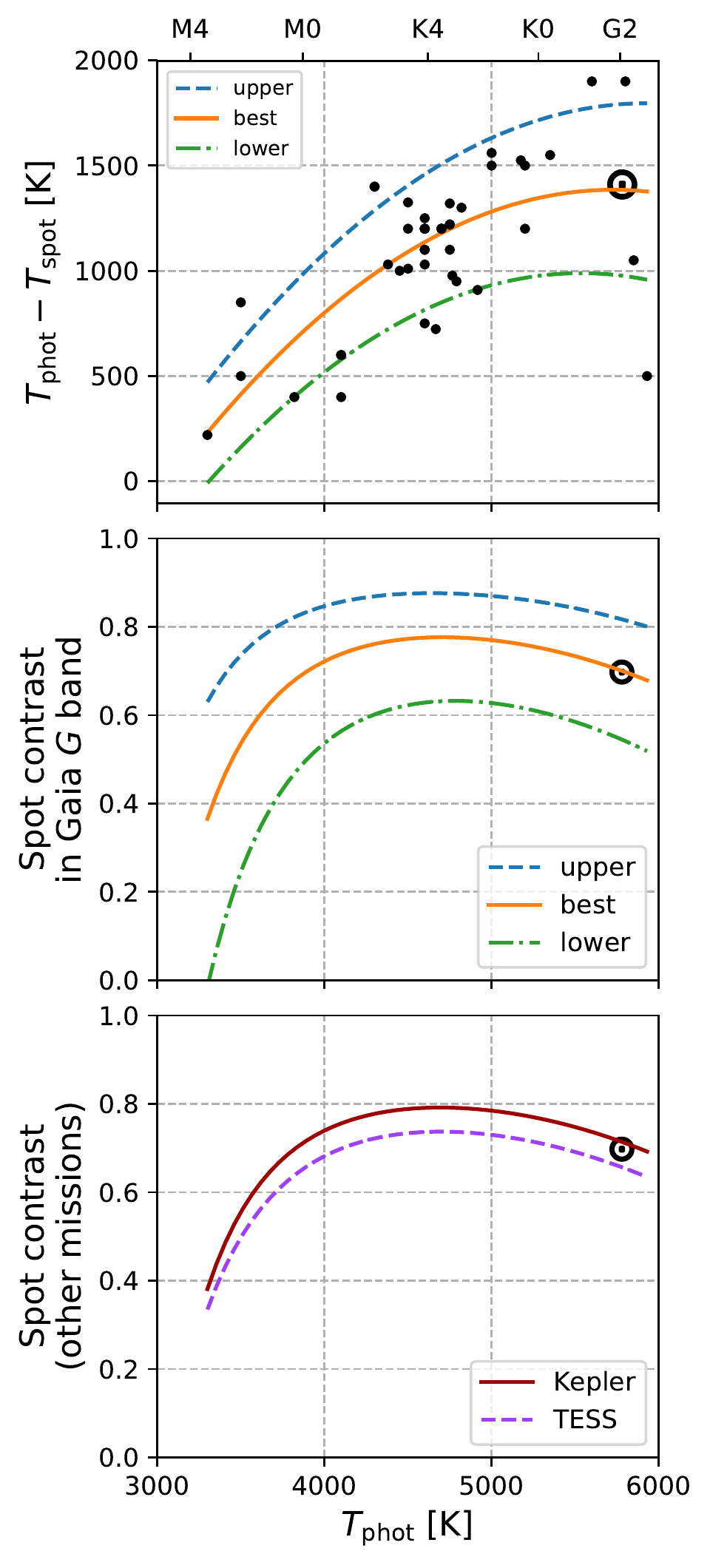}
\caption{{\sl Upper}: Measured temperature difference between the the mean stellar photosphere temperature $T_{\mathrm{phot}}$ and the starspot temperature $T_{\mathrm{spot}}$, as a function of $T_{\mathrm{phot}}$ (black circles), compiled by \citet{Berdyugina2005}. The middle curve labeled ``best'' is a quadratic fit to the spot contrasts. The  ``min,'' and ``max'' curves roughly approximate the lower and upper envelopes of the spot contrast observations. The Sun's contrast is marked with the symbol $\odot$. 
{\sl Middle}: Spot flux contrasts, approximated by integrating blackbody radiance curves with the temperatures of the photosphere and spot, convolved with the Gaia $G$ bandpass. The area-weighted mean sunspot contrast is $c=0.7$, marked with $\odot$. Stars from spectral types M2 to G2 are consistent with $c=0.7\pm0.1$, so we adopt $c=0.7$ for all stars considered in Section~\ref{sec:jitter}.
{\sl Lower}: Starspot flux contrasts, this time integrated over the Kepler (red curve) and TESS (purple dashed curve) bandpasses. Here we show only the contrast curves for the best-fit quadratic spot temperature relation labeled ``best'' in the uppermost panel. Spot contrasts observed with both the \kepler and TESS missions are within 5\% of the Gaia mission spot contrast.
\label{fig:contrast}}
\end{figure}

Since spot flux contrasts vary as a function of spectral type and filter transmittance, we find a relationship between stellar effective temperature and spot contrasts in temperature and flux. In Figure~\ref{fig:contrast}, we show the observed starspot temperature contrasts of 47 stars reported by \citet{Berdyugina2005}, and we fit a quadratic to the spot temperature contrasts as a function of the stellar photosphere effective temperature. We estimate the spot contrast in the Gaia $G$-band by integrating blackbody radiance curves with the temperatures of the photosphere and spot convolved, with the $G$ filter transmittance.

The Sun's spot contrast in the $G$ band, weighting by the relative areas in the penumbrae and umbrae, is $c\sim0.7$. The best-fit quadratic is consistent with $c = 0.7 \pm 0.1$ for stars with spectral types M2 to G2. Given that 73\% of the stars considered in this paper are within that range of spectral types, we choose to use spot contrast $c=0.7$ for all stars (more on the star sample in Section~\ref{sec:jitter}).

\section{Simulating starspot-induced astrometric jitter} \label{sec:sim}

We compute stellar centroid jitter for the starspot distributions of the Sun, GJ 1243 and KIC 7174505. These stars represent different examples of magnetic activity --- the Sun has many small, short-lived spots; GJ 1243 has a few large, long-lived spots; and KIC 7174505 may have extremely large spots. The differences in activity may arise from different dynamo mechanisms for each star, since the Sun has a convective envelope, and GJ 1243 may be fully convective. The configuration of spots on KIC 7174505 and GJ 1243 can be significantly more asymmetric than sunspots, thus producing much larger astrometric signals. In general, the stellar inclination angles for a stars is not known, so we assume that best-case scenario the stars all have stellar inclination $i_s = 90^\circ$, with their rotation axes aligned with the sky plane.

A third star with a well-characterized spot distribution the K4V star HAT-P-11. We do not consider HAT-P-11 because its spectral type and activity is an intermediate case between the Sun and GJ 1243. It has a Sun-like distribution of spots, and spot coverage between that of the Sun and GJ 1243 \citep{Morris2017a, Morris2017b, Davenport2015}. 

\subsection{The Sun} \label{sec:sun}

The Sun will always be the star with the longest-running record of starspot positions and sizes, so we begin by estimating the sunspot-induced astrometric jitter for a distant observer. 

We collect sunspot positions and areas from the Mount Wilson Observatory (MWO) sunspot catalog of \citet{Howard1984}, which spans seven activity cycles from 1917--1985. The sunspot umbral areas are reported in units of solar hemispheres $A_{\mathrm{umb}}$, which are related to the total spot radii $R_{\mathrm{spot}}$ in units of solar radii $R_\odot$ by
\begin{equation}
\frac{R_{\mathrm{spot}}}{R_\odot} = \sqrt{2 (A_{\mathrm{umb}} + A_{\mathrm{pen}})},
\end{equation}
where $A_{\mathrm{pen}}$ is the area in penumbrae. Following \citet{Solanki2003}, we adopt the the penumbral-to-umbral area ratio to be approximately $A_{\mathrm{pen}}/A_{\mathrm{umb}} \sim 4$, so $A_{\mathrm{umb}} + A_{\mathrm{pen}} = 5 A_{\mathrm{umb}}$, and 
\begin{equation}
R_{\mathrm{spot}} \approx \sqrt{10 A_{\mathrm{umb}}} R_\odot.
\end{equation}
Also following \citet{Solanki2003}, we chose the mean flux emitted by a sunspot to be 70\% of the flux of the mean photosphere. 

We define the centroid of the Sun as the flux-weighted mean astrometric position of the Sun in the sky plane for an observer with perfect seeing. We compute the solar centroid by integrating over the Earth-facing solar surface, using all spots reported by \citet{Howard1984}. 

\begin{figure}
\begin{center}
\includegraphics[scale=0.8]{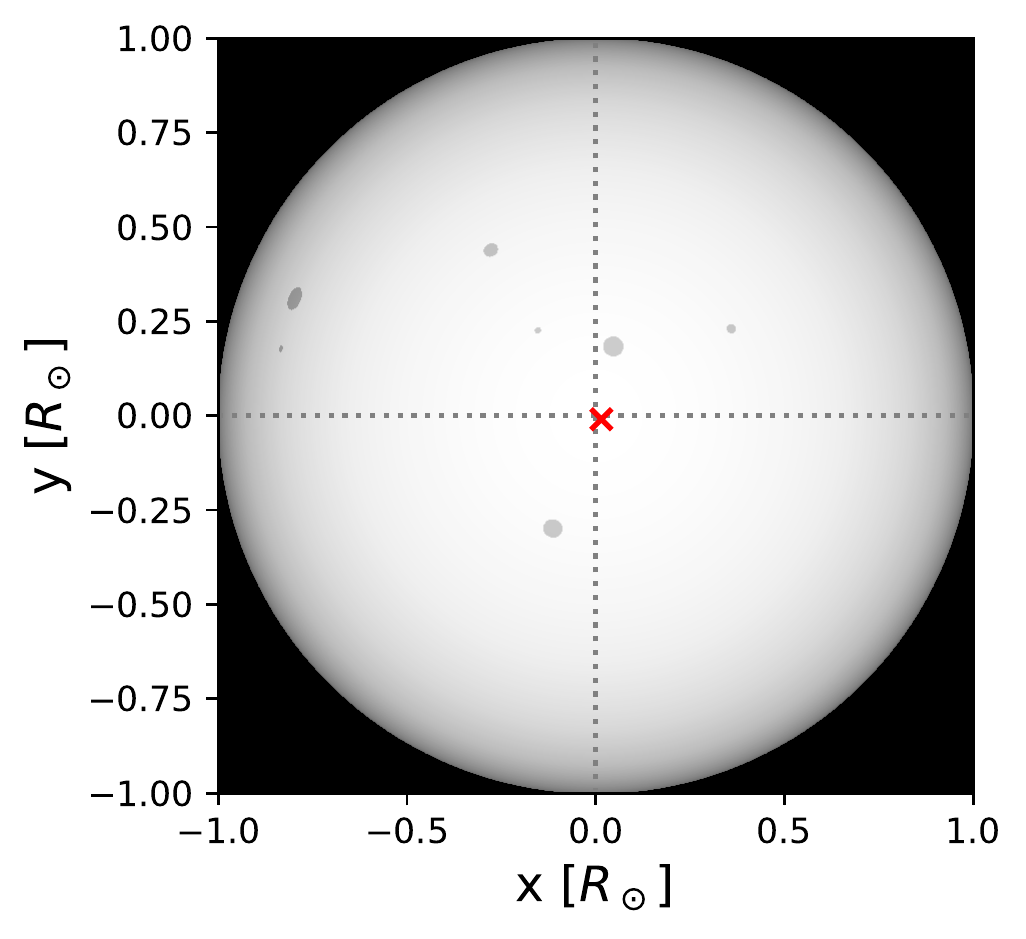}
\end{center}
\caption{Example sunspot record of the Sun on 1959 May 10 UTC. The gray circles represent sunspots, the white disk represents the solar photosphere. The red ``x'' represents the solar centroid in this image, exaggerated by a factor of 100 to make its offset from the origin visible on the diagram.} \label{fig:sunexample}
\end{figure}
A typical entry in the MWO catalog records about three spots in a day, reaching up to 14 spots on the most active day recorded. The median sunspot radius is $R_{\mathrm{spot}} \approx 0.01 R_\odot$. Recorded spot longitudes span the range $[-60^\circ, 51]$ and latitudes span $[-44^\circ, 51^\circ]$. We plot an example sunspot record from 1959 May 10 UTC in Figure~\ref{fig:sunexample}.

The simple sunspot model does not account for faculae, which are bright components of active regions on the Sun. We ignore faculae in our models of the rotationally modulated stellar centroid calculations for two reasons. First, sunspots dominate the Sun's variability in the optical (though the total solar irradiance integrated over all wavelengths is actually greater at active maximum than minimum) \citep{Shapiro2016}. In addition, the impact of faculae on solar brightness modulations would be greatest for an observer viewing the Sun pole-on \citep{Radick1998, Shapiro2016}, but we consider equator-on configurations in this work. 

\begin{figure}
\begin{center}
\includegraphics[scale=0.6]{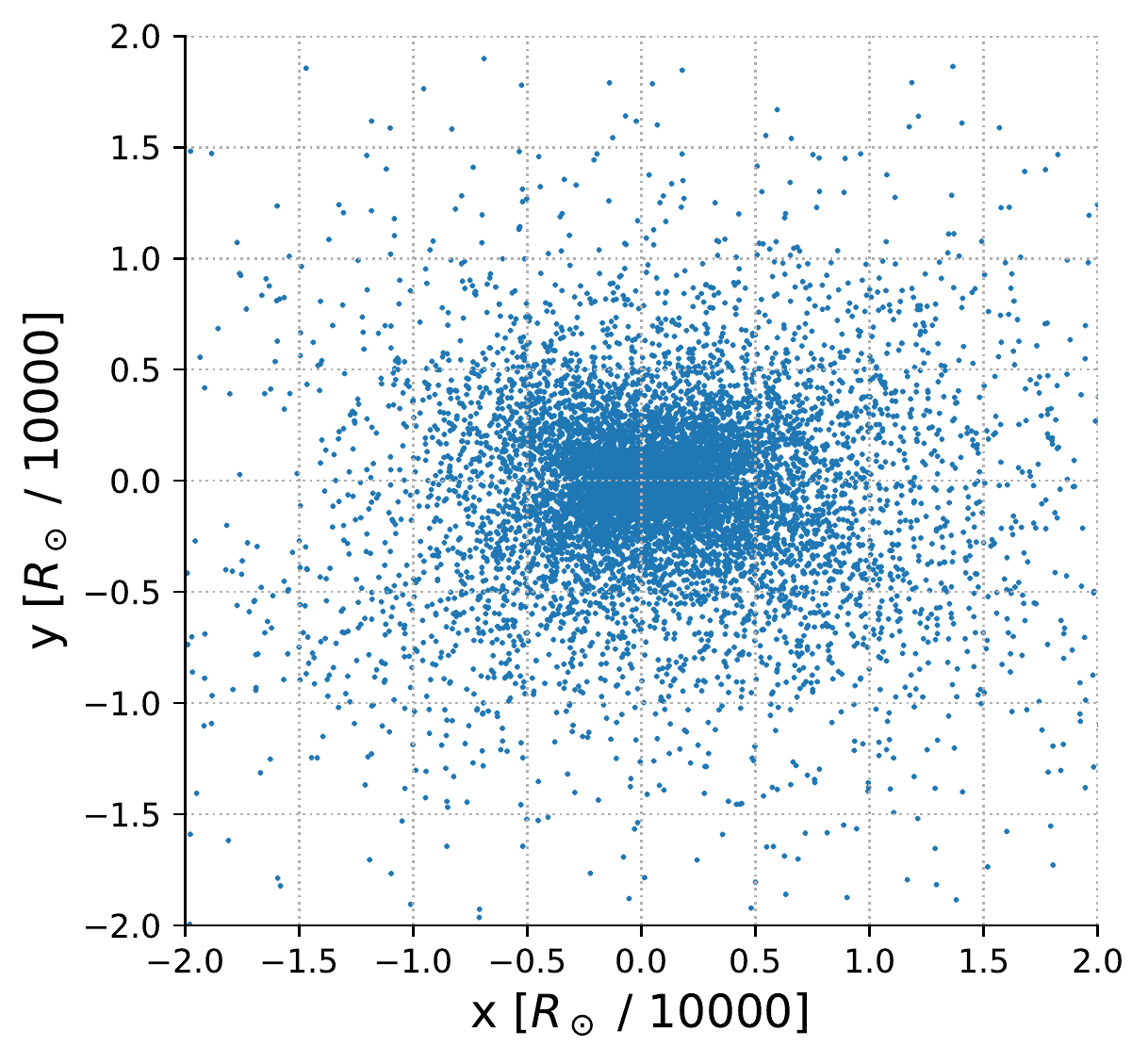}
\includegraphics[scale=0.6]{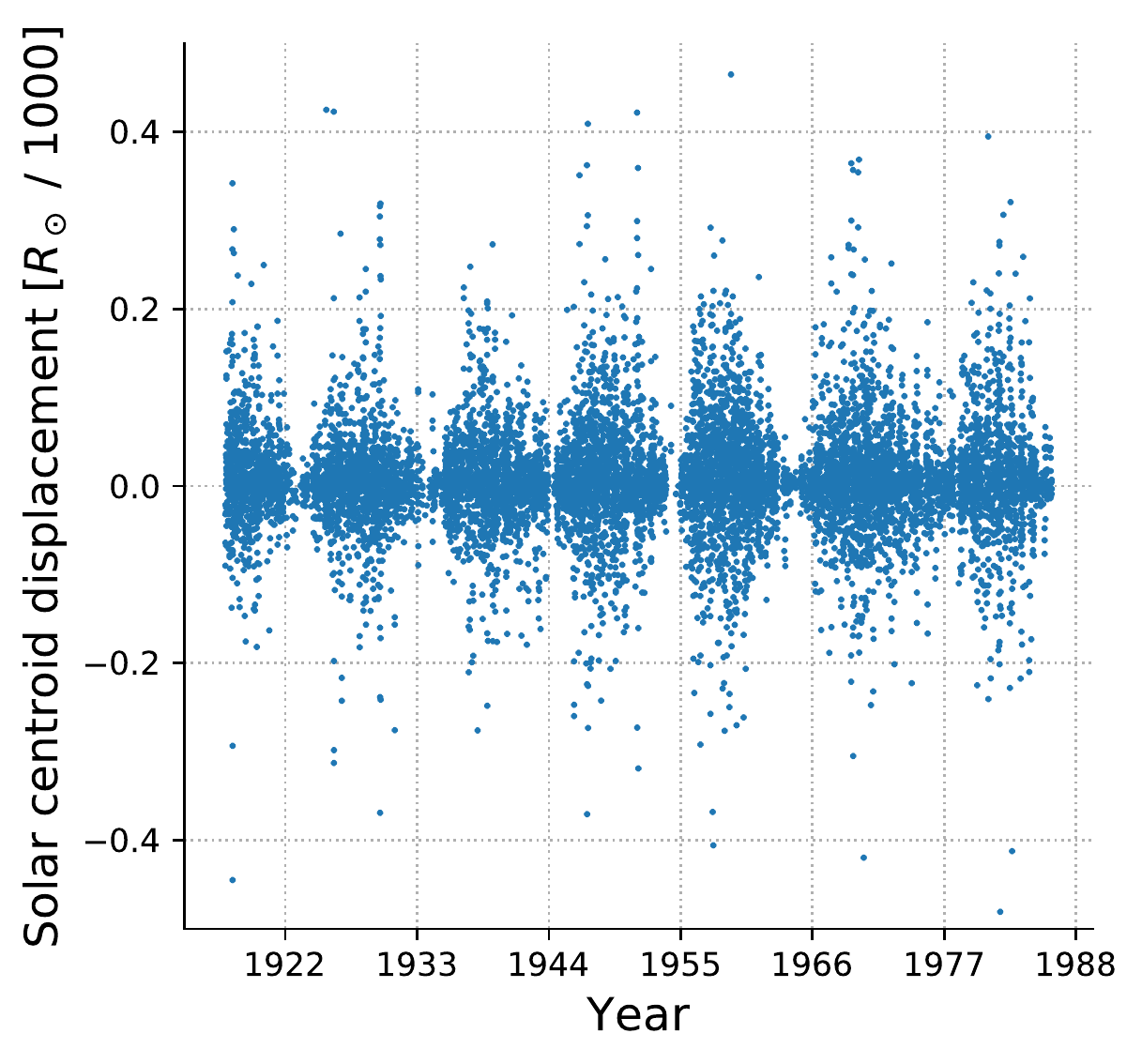}
\end{center}
\caption{\textsl{Upper}: Position of the solar centroid from 1917--1985, accounting only for astrometric shift due to sunspots. The $y$ coordinate is aligned with the solar rotation axis, and the spread in centroids is broader about the equatorial coordinate. \textsl{Lower}: The absolute centroid displacement from the true solar centroid, $r=\sqrt{x^2+y^2}$, throughout seven activity cycles. Near activity maximum, the centroid displacement is greater than at minimum.} \label{fig:centroid}
\end{figure}
The distribution of centroids of the Sun at each date in the \citet{Howard1984} catalog is shown in Figure~\ref{fig:centroid}. The median absolute deviation of the centroids in the $x$ and $y$ directions are 24 and $15 \mu R_\odot$, respectively. 

For comparison, the reflex motion of the Sun about the barycenter due to Jupiter's orbit is roughly the size of the solar radius. The Sun's reflex motion due to the Earth is about $600 \mu R_\odot$, which is still larger than the scale of sunspot-induced astrometric jitter. Once Keplerian orbits are removed from the astrometric measurements of planet-hosting stars, the residual scatter may contain the signals stellar activity discussed here.

\subsection{GJ 1243}

The distribution of sunspots on the solar photosphere is dictated by the physics of the solar dynamo \citep[see reviews by][]{Charbonneau2014, Hathaway2015}. Mid- to late-M dwarfs are expected to have fully-convective envelopes, and thus their dynamo activity must be driven by different physical mechanisms than the Sun's \citep{Morin2010}. The distribution of small starspots on fully-convective stars (as a function of age) is not yet known, in general.

One low mass star with a constrained spot distribution is the M4 dwarf GJ 1243. \citet{Davenport2015} fit the \kepler photometry of GJ 1243 with a spot model, and found that the rotational modulation is consistent with two starspots rotating differentially, with lifetimes of order $\sim$years. The authors estimated that the spots could be as large as $R_{\mathrm{spot}} \sim 0.3 R_\star$ --- significantly larger than the largest sunspots relative to the solar radius. The spot latitudes are degenerate with the stellar inclination, so the precise spot latitudes are unknown. The best-fit spot models of \citet{Davenport2015} prefer one spot near the pole, and one spot at lower latitudes. We note that small spots analogous in scale to sunspots may also be present on GJ 1243, but the rotational modulation was dominated by the two largest spots, so we study the astrometric jitter caused by those  dominant spots \citep{Davenport2015}.

We adopt the low-latitude starspot of GJ 1243 as another prototype for producing astrometric jitter. When compared with the relatively small sunspots, the large, low-latitude spot observed on GJ 1243 can drive significantly more centroid jitter. To construct a best-case scenario for observing spot-induced jitter, we place a spot with radius $R_{\mathrm{spot}} = 0.3R_\star$ on the stellar equator, and view it with stellar inclination $i_s = 90^\circ$. 

We compute the centroid of GJ 1243 throughout a full rotation following the procedure in Section~\ref{sec:sun}. The maximum displacement of the centroid of GJ 1243 is $0.01 R_\star$ --- roughly a factor of 10 greater than the maximum observed solar astrometric jitter. 

\subsection{KIC 7174505} \label{sec:superflarestar}

\begin{figure}
\begin{center}
\includegraphics[scale=0.5]{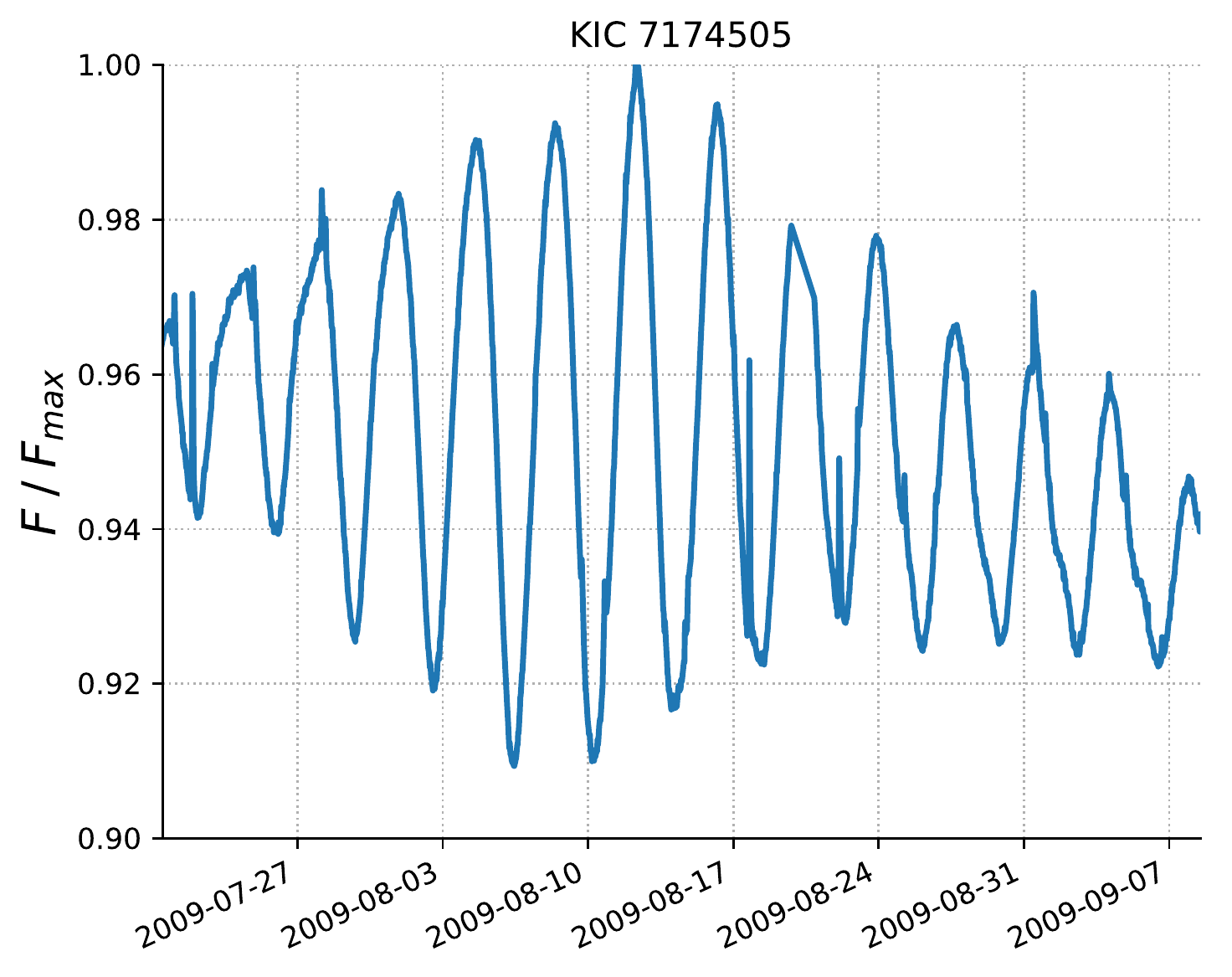}
\end{center}
\caption{A portion of the \kepler light curve of superflare star KIC 7174505, which shows large flux modulation with the rotational period of the star,  $P_{\mathrm{rot}} = 3.832 \pm 0.005$ days \citep{McQuillan2014}.} \label{fig:kic}
\end{figure}

Young G stars in the \kepler field which produce superflares have been the subjects of extensive follow-up observations in the literature \citep[see e.g.:][]{Maehara2012, Maehara2017, Karoff2016, Notsu2013, Notsu2015a, Notsu2015b}. One star with large rotational flux modulation is the superflare star KIC 7174505, with $T_{\mathrm{eff}} \approx 5411$ K and rotation period $P_{\mathrm{rot}} = 3.832 \pm 0.005$ days \citep{Mathur2017, McQuillan2014}. \kepler observed 29 large flare eventson this star, with energies ranging from $7\times10^{34}$ to $2 \times10^{35}$ ergs \citep{Shibayama2013}. A portion of the \kepler light curve of KIC 7174505 is shown in Figure~\ref{fig:kic}.

\citet{Shibayama2013} estimate the spot covering fraction on KIC 7174505 from the rotational modulation of the \kepler light curve. Their estimate of the minimum spotted area is roughly 20\% of the stellar hemisphere, which could be concentrated into one or many spots. If we calculate the radius of a single, circular spot with an area equivalent to 20\% of the stellar hemisphere, we have $R_{\mathrm{spot}}/R_\star = 0.63$. We use this extremely large spot as a limiting case in our calculations for spot-induced astrometric jitter. 

We compute the centroid of KIC 7174505 throughout a full rotation following the procedure in Section~\ref{sec:sun}. The maximum displacement of the centroid of KIC 7174505 is $0.05 R_\star$ --- a factor of five greater than the maximum observed astrometric jitter of GJ 1243. 

\subsection{Expected starspot-induced jitter} \label{sec:jitter}

\begin{figure}
\begin{center}
\includegraphics[scale=0.65]{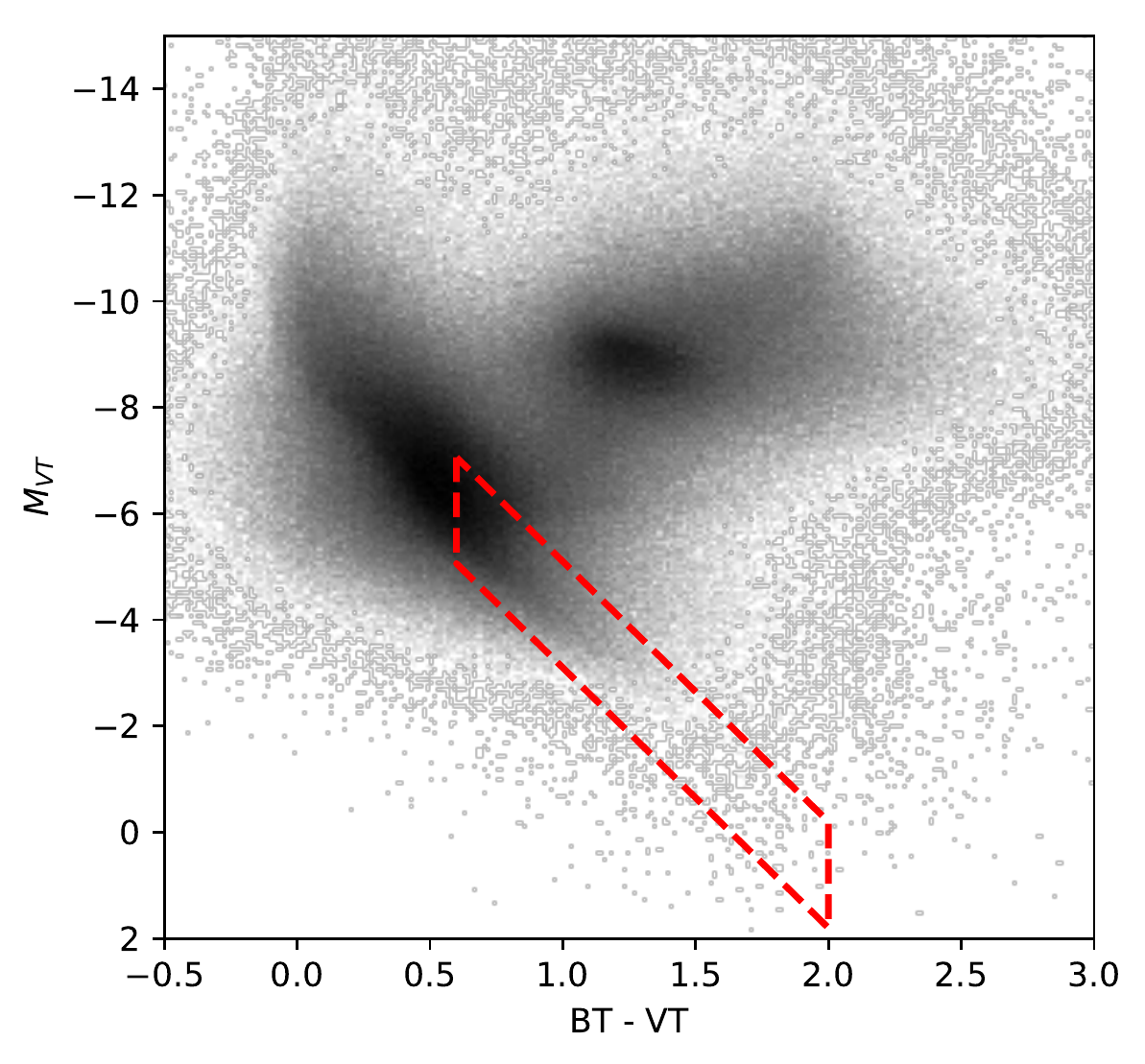}
\end{center}
\caption{Color-magnitude diagram for stars brighter than $G < 12$ in the TGAS catalog, where the shading is the logarithm of density of stars in each pixel. $BT$ and $VT$ are the Tycho-2 bandpasses,  similar to Johnson-Cousins $B$ and $V$. We select only stars within the red polygon, which are likely main sequence stars of approximately solar spectral type or later.} \label{fig:hr}
\end{figure}

We calculate the expected starspot-induced jitter for bright, cool, main sequence stars that will be observed by Gaia. To select stars meeting these criteria, we choose stars with Gaia magnitude $G < 7$ from the Tycho-Gaia Astrometric Solution (TGAS) catalog \citep{Michalik2015}. TGAS combines early Gaia astrometry with measurements from the Tycho-2 astrometric mission to solve for parallaxes of millions of nearby stars. We use Tycho photometry in the VT and BT bands, and the combined Tycho-2/Gaia parallaxes to construct a color-magnitude diagram of the bright TGAS stars. We narrow our sample to stars on the main sequence with colors $0.6 < B - V < 2$. These 8,896 stars are highlighted in red on the color-magnitude diagram in Figure~\ref{fig:hr}. It is likely that there are binaries in this sample of stars. We include all stars in our analysis until reporting the best-case targets in Table~\ref{tab:stars}, which have been filtered to remove known binaries.

The TGAS sample does not include the brightest stars in the sky \citep{Michalik2015}. Ongoing efforts throughout the beginning of the Gaia mission have made progress towards measuring astrometry of naked-eye stars, and it is possible that ultimately there will be no bright limit for Gaia astrometry \citep{Martin-Fleitas2014, Sahlmann2016}. Since the final mission bright limit is not yet established, we choose to consider only the stars with Gaia astrometry already published in TGAS, and ignore the brighter stars which may be accessible to Gaia astrometry in the future.  

We calculate Gaia's expected end-of-mission astrometric precision, $\sigma_{\mathrm{Gaia}}$, for the stars in the bright sample. \citet{Perryman2014} outline an algorithm for calculating the expected Gaia astrometric precision for any star, which we reproduce here. The astrometric signal-to-noise achieved in a single Gaia visit $\sigma_{\mathrm{fov}}$ can be computed with:
\begin{equation}
\sigma_{\mathrm{fov}}^2 = \sigma_\eta^2/9 + \sigma_{\mathrm{att}}^2 + \sigma_{\mathrm{cal}}^2
\end{equation}
where the contributions from attitude errors $\sigma_{\mathrm{att}}$ and calibration errors $\sigma_{\mathrm{cal}}$ are both approximately $\sigma_{\mathrm{att}} \approx \sigma_{\mathrm{cal}} = 20 \,\mathrm{\mu arcseconds}$, and the centroiding error $\sigma_\eta$ for each of Gaia's nine CCDs is	
\begin{equation}
\sigma_\eta^2 = 53000 z + 310 z^2.
\end{equation}
$z$ is a function of the inverse number of photons in the image, which is a function of a star's $G$ magnitude,
\begin{equation}
z = 10^{0.4\left(\max[G, 12] - 15\right)}. \label{eqn:z}
\end{equation}
The number of expected visits at a given star over the nominal five-year mission $N^\prime_{\mathrm{fov}}$ is a function of the galactic latitude $b$ of the star (see \citealt{Perryman2014} Table 1 for $N^\prime_{\mathrm{fov}}$ estimates at each $b$). The maximum number of photons received from stars brighter than $G < 12$ is capped by CCD gating, which prevents saturation \citep{Perryman2014}. One consequence of this observing strategy is that the astrometric precision is roughly uniform for all targets brighter than $G < 12$. 

Thus, we can approximate the Gaia end-of-mission astrometric precision for each bright star in the sample with 
\begin{equation}
\sigma_{\mathrm{Gaia}}(G, b) = \sigma_{\mathrm{fov}}(G) / \sqrt{N^\prime_{\mathrm{fov}}(b)}. 
\end{equation}
In other words, the distribution of the Gaia-measured centroids of a single, inactive star without companions will have RMS width $\sigma_{\mathrm{Gaia}}$. Barring unanticipated systematics, scatter measured in the centroid of a star exceeding $\sigma_{\mathrm{Gaia}}$ may be interpreted as the signature of stellar multiplicity, massive planets, or starspots.

\subsection{Sun-like distribution of starspots} \label{sec:sunspot_jitter}

We calculate the spot-induced centroid jitter for the bright TGAS main sequence stars, assuming a Sun-like distribution of spots. We convert the centroid jitter observed on the Sun, in units of $R_\odot$, to the expected RMS angular astrometric jitter $\sigma_{\mathrm{jitter}}$ by observing the solar centroid offsets at the distance of each star in the TGAS sample. Then we normalize the observed jitter by the expected Gaia astrometric uncertainty. 

The expected astrometric jitter produced by a Sun-like distribution of starspots on the stars in the bright TGAS main sequence sample is shown in Figure~\ref{fig:jitter}. The spot-induced jitter $\sigma_{\mathrm{jitter}}$ is normalized by the cumulative, end-of-mission astrometric uncertainty of Gaia, after considering the single-measurement astrometric precision on that star, and the total number of measurements during the nominal five-year mission. Even for the brightest star in the sample, the K0V star $\sigma$ Draconis ($G = 4.71$), the spot-induced jitter is only 1\% of the expected astrometric uncertainty. Measurements of astrometric starspot jitter from Sun-like spot distributions are thus beyond the reach of the nominal five-year Gaia mission.


\subsection{GJ 1243-like distribution of starspots} \label{sec:gj1243_jitter}


We calculate the spot-induced centroid jitter for the bright TGAS main sequence stars, assuming a GJ 1243-like distribution of spots. We convert the centroid jitter observed on GJ 1243, in units of $R_\star$, to the expected RMS angular astrometric jitter $\sigma_{\mathrm{jitter}}$ by observing the stellar centroid offsets at the distance of each star in the TGAS sample. We estimate stellar radii for each star using the color--radius relations of \citet{boyajian2012}, and for a small subset of the stars, we combine the interferometric stellar angular diameters from \citet{boyajian2012} with TGAS parallaxes \citep{Michalik2015}. Then we normalize the observed jitter by the expected Gaia astrometric uncertainty. 

The expected astrometric jitter produced by a GJ 1243-like distribution of starspots on the stars in the bright TGAS main sequence sample is shown in Figure~\ref{fig:jitter}. The starspot jitter is comparable or less than the predicted astrometric uncertainty for all stars considered here. At the end of the nominal five-year mission, Gaia astrometry could not detect GJ 1243-like activity on these stars.

\subsection{KIC 7174505-like distribution of starspots} \label{sec:kic_jitter}

\begin{figure}
\begin{center}
\includegraphics[scale=0.6]{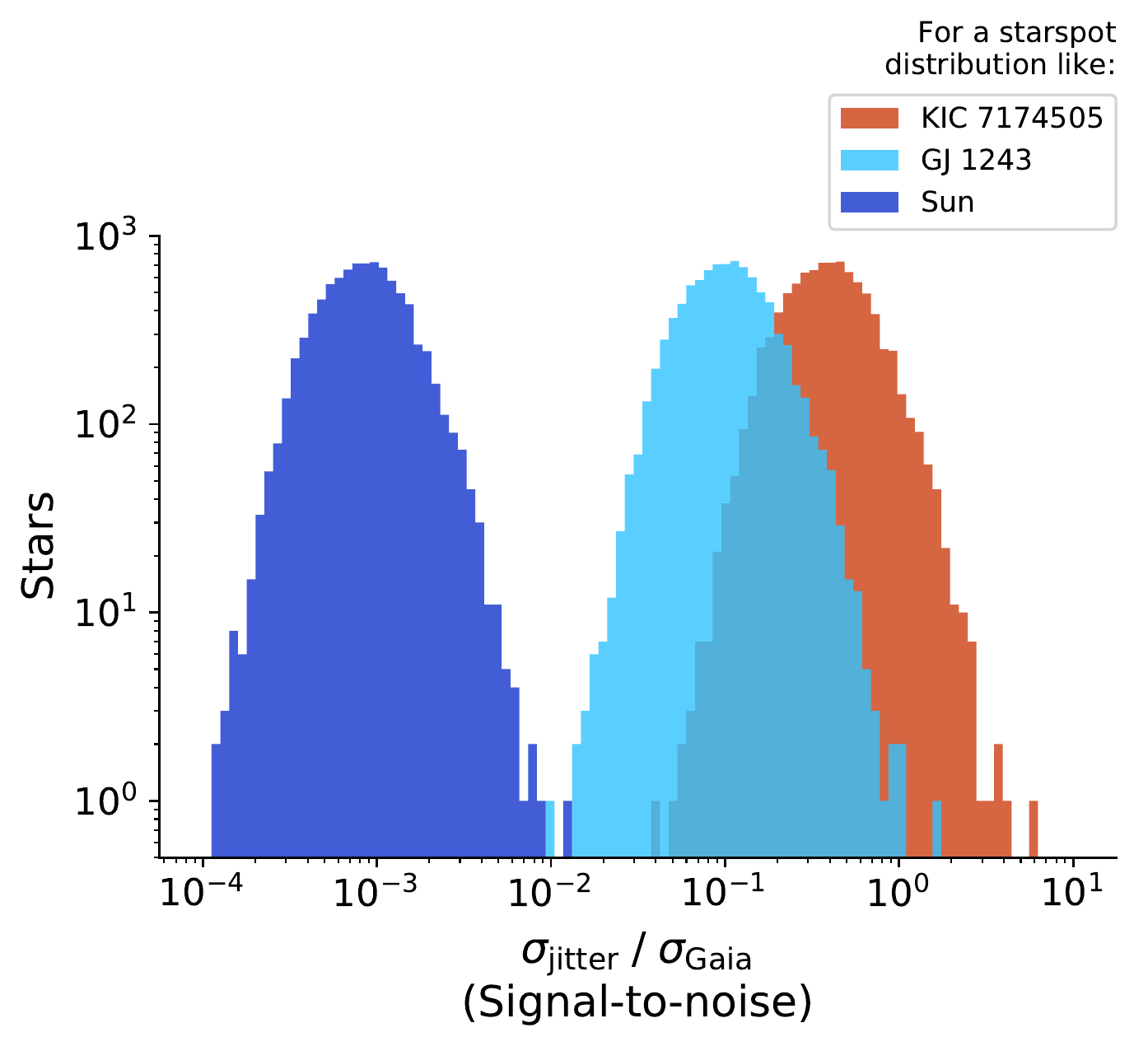}
\end{center}
\caption{Maximum expected starspot-induced jitter for 8,896 nearby Gaia stars in TGAS, normalized by the estimated astrometric uncertainty for a nominal 5 year mission, assuming starspot distributions like the Sun, GJ 1243, and KIC 7174505. These maximum jitter estimates only consider stellar inclination $i_s = 90^\circ$. Details for the eight best non-binary stars are enumerated in Table~\ref{tab:stars}.} \label{fig:jitter}
\end{figure}

We calculate the spot-induced centroid jitter for the bright TGAS main sequence stars with a distribution of starspots like KIC 7174505 following the procedure outlined in Section~\ref{sec:gj1243_jitter}. The expected astrometric jitter is shown in Figure~\ref{fig:jitter}. The starspot jitter is larger than the predicted astrometric uncertainty by a factor of three for six stars in the sample, and the jitter exceeds the noise by a factor of six for one non-binary star: AX Mic. We discuss these stars in more detail in Section~\ref{sec:targets}. Gaia astrometry can detect the most extreme centroid offsets even in a fraction of the nominal five-year mission.

\subsection{Comparison with other works}

\citet{Eriksson2007} and \citet{Catanzarite2008} devise starspot jitter models based on observations of the Sun and other stars. Both groups report a maximum centroid jitter of about $\sim 2 \,\mu$AU for solar-like activity, which is a bit smaller than our maximum estimate of the solar centroid offset based on the observations of \citet{Howard1984} of $\sim 3 \,\mu$AU. \citet{Eriksson2007} also estimate the centroid offsets for M2--M9V type stars in the range $0.4-10 \,\mu$AU, which is smaller than our expectation for the jitter from a GJ1243-like spot distribution, which contributes $\sigma_{\mathrm{jitter}} \approx 35\, \mu$AU. We are unable to find analogous estimates in the literature of starspot jitter from superflare stars like KIC 7174505, which can produce apparent jitter of $100\,\mu$AU.

\subsection{Activity cycles}

The number of sunspots and their latitude distribution changes throughout the solar activity cycle. Near activity minimum, there are sometimes no spots on the Sun at all, and near maximum there can be more than ten. A distant observer could measure centroid offsets due to starspots near activity maximum, but the jitter would diminish at activity minimum -- Figure~\ref{fig:centroid} shows the varying scatter in the solar centroid throughout seven solar activity cycles based on real solar observations. 

We showed in Section \ref{sec:sim} that sunspots would be undetectable by astrometric jitter, jitter from GJ 1243-like spots may be detectable, and jitter from KIC 7174505 would be detected handily for the nearest stars. Is the astrometric signal-to-noise and sampling in time sufficient to measure changes in astrometric jitter throughout stellar activity cycles? 

Here we consider the best-case scenario for detecting activity cycles on stars with spots like KIC 7174505. We seek to determine the difference in astrometric jitter near activity minimum, when $\sigma_{\mathrm{jitter}} \rightarrow 0$, and near activity maximum, when we assume the maximum jitter is achieved by the starspot distribution observed on KIC 7174505. 

Suppose KIC 7174505-like stars have an activity cycle period $P_{\mathrm{cyc}} \sim 10$ years, and spend five years near maximum and five years near minimum. If the Gaia extended mission observed that star over an extended mission lasting 10 years, it will measure the stellar centroid $N^\prime_{\mathrm{fov}}$ times near activity minimum, when the astrometric centroid scatter will be entirely contributed by random and systematic errors. We simulate these observations by drawing $N^\prime_{\mathrm{fov}}$ samples from the distribution $\mathcal{N}(0, \sigma_{\mathrm{fov}}^2)$. Then we observe each star another $N^\prime_{\mathrm{fov}}$ times near activity maximum during a five-year extended mission, when the astrometric scatter will be the quadrature sum of the random errors $\sigma_{\mathrm{fov}}$ and the spot-induced jitter $\sigma_{\mathrm{jitter}}$. We simulate these observations by drawing $N^\prime_{\mathrm{fov}}$ samples from the distribution $\mathcal{N}(0, \sigma_{\mathrm{fov}}^2 + \sigma_{\mathrm{jitter}}^2)$. We compute the confidence of the cycle detection measurement with the two-sample Anderson-Darling and Komolgorov-Smirnov (KS) tests to measure the difference in centroid distributions observed at activity maximum and minimum.

In Table~\ref{tab:stars} we list the statistical significance of the difference between the distribution of centroids near active maximum and minimum, for the bright TGAS sample in an extended 10 year Gaia mission. The stars considered show insignificant detections of an activity cycle in this highly-optimized scenario for detecting activity cycles through Gaia astrometry, where we assumed a highly asymmetric spot distribution and an ideal stellar inclination of $i_s = 90^\circ$. We comment on the properties of the star with the best detection of any activity, AX Mic, in the next section.

The activity cycle detection confidences in Table~\ref{tab:stars} assume that we can unambiguously split the astrometric measurements into maximum and minimum activity subsets. In practice, we will not know the phase of each star's activity cycle \textit{a priori}. However, Gaia radial velocity spectra of the calcium infrared triplet may allow us to independently identify phases in activity cycles. The calcium infrared triplet is a calibrated indicator of chromospheric activity \citep[see e.g.:][]{Chmielewski2000, Cenarro2001, Cauzzi2008, Zerjal2013, BoroSaikia2016, Robertson2016, Martin2017}. True cycles in astrometric jitter should be accompanied by increased emission in the infrared triplet \citep{Andretta2005, Busa2007}. A possible third independent measurement of activity cycles from Gaia is the time-series photometry it will collect, which may yield long-baseline photometry similar to the \kepler full-frame images, which \citet{montet2017} used to search for activity cycles in \kepler stars.

\subsection{Sensitivity to spot coverage}

It is common to use stellar variability as an estimate of the spot coverage
of stars;  this is adequate for stars that are affected by a single spot,
or two anti-podal spots since only one spot is present on the visible disk
at a time.  Unfortunately, more than two spots that are distributed 
in longitude
cause a weaker amplitude variability:  as one spot rotates off of the
visible disk of the star, another rotates onto it, causing a reduction in
the overall amplitude of variability.  A similar effect has been pointed
out in the context of planetary mapping:  the odd spherical harmonics
induce no change in brightness \citep{Cowan2008}.

\begin{figure}
    \centering
    \includegraphics[scale=0.38]{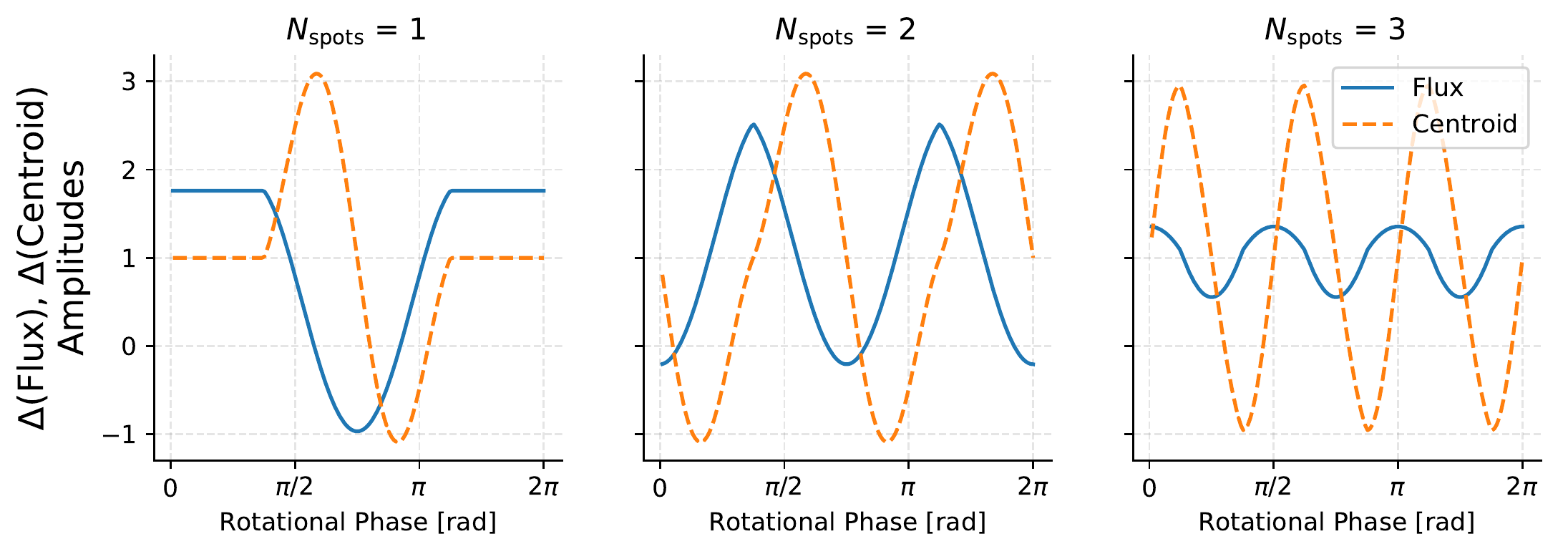}
    \caption{Relative amplitude of variability for one (left),
    two (middle) and three (right) spots of equal radii, in flux (blue)
    and astrometric angle (orange dashed).}
    \label{fig:photo_vs_astro}
\end{figure}

In particular, a distribution of three spots with 120$^\circ$ separation
can lead to a very small amplitude of variability.  However, the
astrometric signal of three spots can still be strong despite weak
flux variability. Figure~\ref{fig:photo_vs_astro} shows the relative 
amplitude of variability of the total flux for one, two and three equatorial
spots, distributed evenly in longitude, and viewed at $i_s = 90^\circ$.  In the one- and two- spot cases, the relative
amplitude of variability is similar for photometry and astrometry.
However, in the three-spot case, the photometric variability is
suppressed by a factor of 6.5 relative to the astrometric variability.

\section{Best targets for activity detection with Gaia} \label{sec:targets}

We evaluate the astrometric signal-to-noise expected from the nominal five year Gaia mission for the brightest, nearest main sequence stars in the TGAS sample, and list the results for the eight best candidates in Table~\ref{tab:stars}. The last column of the table denotes the expected scale of starspot jitter normalized by Gaia's astrometric precision $\sigma_{\mathrm{jitter}}/\sigma_{\mathrm{Gaia}}$, in the extreme case of a KIC 7174505-like starspot configuration, observed with stellar inclination $i_s = 90^\circ$. This should be treated as an upper-limit on the astrometric signal that starspots could produce on main sequence stars. In the sections that follow, we examine measurements of each star's activity in the literature.

\begin{table*}
\begin{tabular}{lccccccccccc}
Star & HIP & Spectral & $R_\star$ & $G$ & Distance & $N^\prime_{\mathrm{fov}}$ & $\sigma_{\mathrm{jitter}}$ & $\sigma_{\mathrm{jitter}}/\sigma_{\mathrm{Gaia}}$ & \multicolumn{2}{c}{Best-case cycle sig.:} \\ 
 &  & Type & [$R_\odot$] & [mag] & [$\mathrm{pc}$] & & [$\mathrm{\mu as}$] & (S/N) & KS & Anderson\\ \hline\hline
AX Mic & 105090 & M1V & 0.589\tablenotemark{a} & 5.881 & 3.98 & 107.4 & 20.0 & 6.1 & 0.45 & 0.37 \\
$\sigma$ Dra & 96100 & K0V & 0.789\tablenotemark{a} & 4.711 & 5.76 & 58.5 & 18.5 & 4.1 & 0.5 & 0.43 \\
GX And & 1475 & M2V & 0.459\tablenotemark{b} & 7.096 & 3.56 & 55.6 & 17.4 & 3.8 & 0.48 & 0.43 \\
HD 79211 & 120005 & M0V & 0.572\tablenotemark{b} & 6.948 & 6.29 & 107.4 & 12.3 & 3.7 & 0.51 & 0.45 \\
LHS 3531 & 99701 & M0V & 0.564\tablenotemark{a} & 7.134 & 6.16 & 69.1 & 12.4 & 3.0 & 0.51 & 0.45 \\
HD 222237 & 116745 & K3+V & 0.772\tablenotemark{a} & 6.669 & 11.39 & 107.4 & 9.2 & 2.8 & 0.51 & 0.47 \\
HD 36395 & 25878 & M1.5Ve & 0.531\tablenotemark{b} & 6.986 & 5.65 & 55.6 & 12.7 & 2.8 & 0.53 & 0.48 \\
Gl 625 & 80459 & M1.5V & 0.431\tablenotemark{a} & 8.995 & 6.49 & 107.4 & 9.0 & 2.7 & 0.5 & 0.46 \\
\end{tabular}
\caption{Expected signal-to-noise for the astrometric jitter due to KIC 7174505-like starspots for the best non-binary targets. The columns are: the Hipparcos number, spectral type, Gaia $G$ band magnitude, distance computed from the TGAS parallax, anticipated number of Gaia visits to that particular star $N^\prime_{\mathrm{fov}}$, astrometric jitter from KIC 7174505-like starspots $\sigma_{\mathrm{jitter}}$,  the ratio of the starspot-induced astrometric jitter to the expected astrometric precision over five years of Gaia observations $\sigma_{\mathrm{jitter}}/\sigma_{\mathrm{Gaia}}$. The astrometric precision the targets above (and for Gaia targets brighter than $G < 12$) is constant, $\sigma_{\mathrm{fov}} = 34.2 \,\mu$arcseconds. The last two columns list the expected statistical significance of the astrometric detection of stellar activity cycles for each star over a 10 year extended mission. Radii marked \tablenotemark{a} are estimates from color--radius relation of \citet{boyajian2012}; radii marked \tablenotemark{b} are computed from interferometry by \citet{boyajian2012}, combined with parallaxes from TGAS \citep{Michalik2015}.
\label{tab:stars}}
\end{table*}

\subsection{AX Mic (HD 202560, Gl 825)}

AX Microscopii is a single M1V flare star, with a TGAS distance of 3.9 pc \citep{Young1987, Byrne1981}. \citet{Byrne1989} measured the X-ray flux of AX Mic, and found that its transition region emission is similar to the Sun's, despite being significantly smaller than the Sun. The authors suggest that AX Mic's strong X-ray emission per unit surface area could be achieved with a few active regions on the stellar surface with a filling factor of 0.02\% --- similar to the typical solar spot filling factor.  \citet{Isaacson2010} measured small variations in the chromospheric emission of AX Mic within the range $1.223 < S < 1.386$ with Keck/HIRES.

\subsection{$\sigma$ Draconis (HD 185144)}

$\sigma$ Draconis is a K0V star which was extensively monitored by the Lick-Carnegie Exoplanet Survey Team (LCES) HIRES/Keck Precision Radial Velocity Exoplanet Survey \citep{Butler2017}. The 784 spectra of $\sigma$ Dra reveal an apparent magnetic activity cycle with period $P_{\mathrm{cyc}} \approx 6$ years. 

\subsection{GX And (HD 1326, Gl 15 A)}

GX Andromedae is a M2V flare star at a TGAS distance of 3.6 pc, which hosts at least one exoplanet, usually referred to as Gl 15 A b. The mean chromospheric emission from GX And is $\left< S \right> = 0.524$, and $S$ varies on two timescales. The shorter timescale is likely the stellar rotation period at $P_{\mathrm{rot}} = 44$ d, and the authors suggest that the longer periodicity is a magnetic activity cycle, with period $P_{\mathrm{cyc}} = 9 \pm 2.5$ yr. 

\subsection{HD 79211 (Gl 338 B)}

HD 79211 is a M0Ve flare star. \citet{Pace2013} report significantly variable chromospheric emission on the range $1.55 < S < 2.02$, which is in good agreement with the range observed by \citet{Isaacson2010}, from $1.556 < S < 1.961$. If the variations in chromospheric emission are driven by rotational modulation from active regions, there may be spot-induced jitter to observe in the Gaia astrometry. 

\subsection{HD 36395 (Gl 205)}

HD 36395 is a weakly active M1V star. \citet{Hebrard2016} collected Zeeman-Doppler imaging of HD 36395, and measured the rotation period $P_{\mathrm{rot}} = 33.63 \pm 0.37$ d. The authors found that the star's large-scale magnetic field is mostly poloidal, and the radial velocity periodogram showed signal at multiple periods, separated by about 10 d. They suggest that the multiple periods could arise from dark spots rotating differentially -- one near the pole and one nearer to the equator, in a configuration reminiscent of GJ 1243. 

\subsection{Gl 625 (G 202-48)}

Gl 625 is an M2V star with one known planet --- a super-Earth which orbits at the inner edge of the stellar habitable zone \citep{Mascareno2017}. There are several signals in the periodogram of the $S$-index of Gl 625, which may be the signature of an activity cycle with period $P_{\mathrm{cyc}} = 3.4$ yr. 

\begin{figure}
\centering
\includegraphics[scale=0.6]{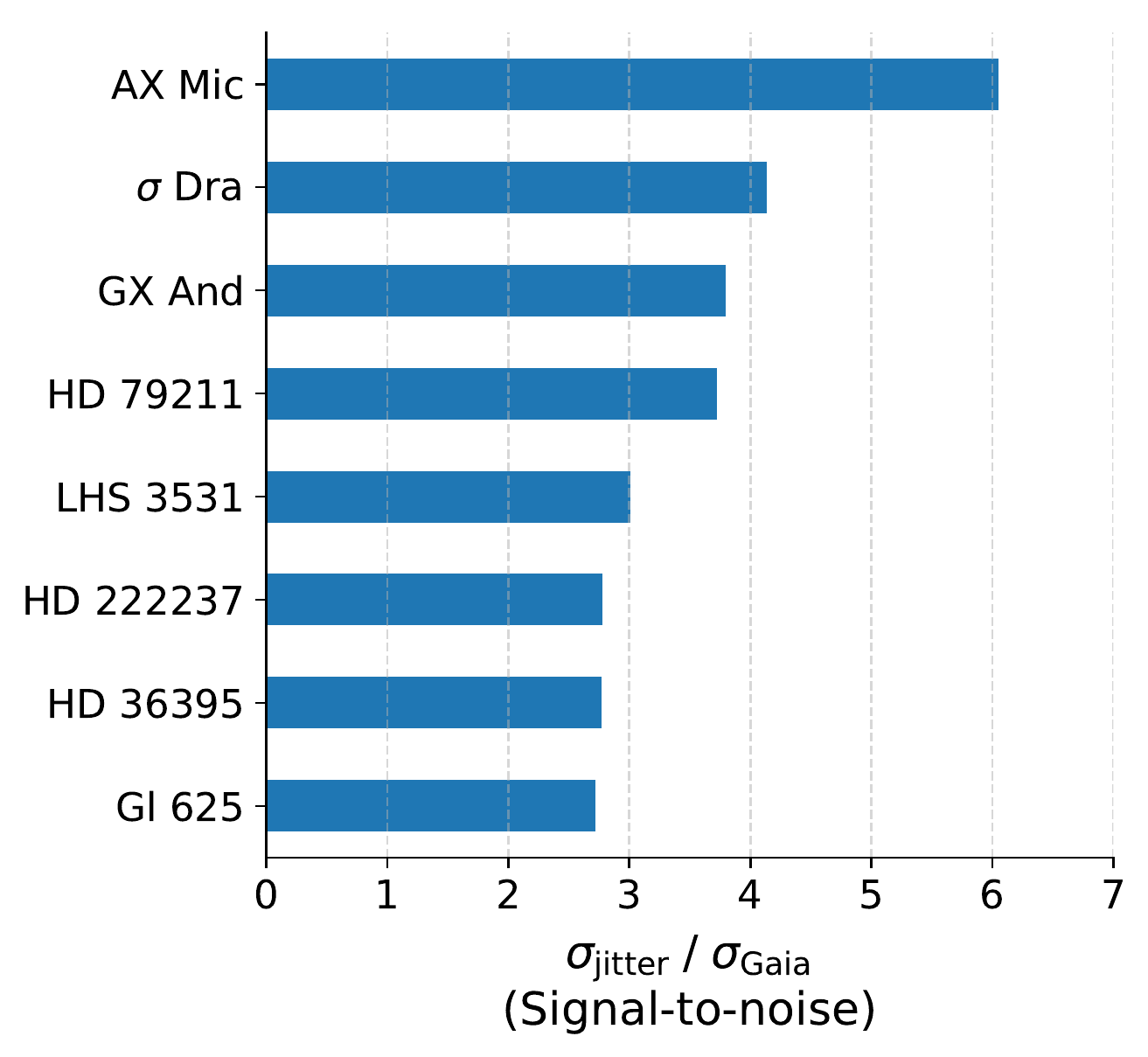}
\caption{The ratio of the starspot-induced astrometric jitter to the expected astrometric precision over five years of Gaia observations $\sigma_{\mathrm{jitter}}/\sigma_{\mathrm{Gaia}}$ for the targets in Table~\ref{tab:stars}. \label{fig:best}}
\end{figure}

\section{Conclusion and Discussion}

The ESA Gaia mission will measure astrometry of bright nearby stars with unprecedented precision. We show that Gaia's astrometric precision will be sufficient to measure the astrometric shifts in the centroids of stars due to dark starspots on some of the nearest low-mass stars, which were included in the TGAS sample. The best candidate for the detection of astrometric jitter induced by starspots is AX Mic, an active M1V flare star. In the most optimistic scenario --- where each star hosts large KIC 7174505-like spots which vary on a 10 year activity cycle --- the variation in astrometric jitter throughout the stellar activity cycles is not detectable with Gaia astrometry alone. The top three targets for astrometric detections of activity (included in the TGAS sample) are: AX Mic, $\sigma$ Dra, and GX And. Upcoming efforts to measure astrometry of the brightest stars with Gaia may expand the sample of stars with detectable activity cycles \citep{Martin-Fleitas2014, Sahlmann2016}.

Other possible applications of astrometric spot variability detection
may include: 1). estimating the total spot coverage since astrometry
is relatively more sensitive to longitudinally-distributed spots compared
total flux (which can affect transit transmission spectra for stars
that host planets); 2). deriving the sky-projected angle of inclination
of stars, which may help to constrain alignment of spin axes of stars; 
3). constraining the 3-D inclination of stars; 4). constraining the North-South asymmetry of spot distributions.  We leave
these ideas for future investigation.


\section*{Acknowledgements}

The manuscript benefited from helpful discussions with Andrew Vanderburg. We gratefully acknowledge the software that made this work possible, including: \texttt{ipython} \citep{ipython}, \texttt{numpy} \citep{VanDerWalt2011}, \texttt{scipy} \citep{scipy},  \texttt{matplotlib} \citep{matplotlib}, \texttt{astropy} \citep{Astropy2013}. 

This work has made use of data from the European Space Agency (ESA)
mission {\it Gaia} (\url{https://www.cosmos.esa.int/gaia}), processed by
the {\it Gaia} Data Processing and Analysis Consortium (DPAC,
\url{https://www.cosmos.esa.int/web/gaia/dpac/consortium}). Funding
for the DPAC has been provided by national institutions, in particular
the institutions participating in the {\it Gaia} Multilateral Agreement. 

This research has made use of the SVO Filter Profile Service (\url{http://svo2.cab.inta-csic.es/theory/fps/}) supported from the Spanish MINECO through grant AyA2014-55216 \citep{rodrigo2012}.

This research has made use of NASA's Astrophysics Data System. This research has made use of the SIMBAD database, operated at CDS, Strasbourg, France \citep{Wenger2000}. This research has made use of the VizieR catalogue access tool, CDS, Strasbourg, France \citep{Ochsenbein2000}. This research made use of the cross-match service provided by CDS, Strasbourg.

\bibliographystyle{mnras}

\bsp	
\label{lastpage}
\end{document}